\documentclass[aps,prl,preprint,superscriptaddress]{revtex4-1}

\usepackage[utf8]{inputenc}
\usepackage{amsmath}
\usepackage{textcomp}
\usepackage{graphicx}
\usepackage{array}
\usepackage{color}
\usepackage{bm}
\usepackage{amssymb}
\usepackage{soul}

\usepackage{titlesec}

\newcommand*\rfrac[2]{{}^{#1}\!/_{#2}}

\titleformat*{\section}{\Large\bfseries}

\begin{document}

\title{Magnetic hyperbolic optical metamaterials}

\author{Sergey~S.~Kruk}
\affiliation{Nonlinear Physics Center and Center for Ultrahigh Bandwidth Devices for Optical Systems (CUDOS),The Australian National University, Canberra, ACT 0200, Australia}

\author{Zi~Jing~Wong}
\affiliation{NSF Nanoscale Science and Engineering Center, University of California, Berkeley, CA 94720, USA}

\author{Ekaterina~Pshenay-Severin}
\affiliation{Nonlinear Physics Center and Center for Ultrahigh Bandwidth Devices for Optical Systems (CUDOS),The Australian National University, Canberra, ACT 0200, Australia}
\affiliation{Institute of Applied Physics, Abbe Center of Photonics, Friedrich-Schiller-Universit\"at Jena, 07743 Jena, Germany}

\author{Kevin~O'Brien}
\affiliation{NSF Nanoscale Science and Engineering Center, University of California, Berkeley, CA 94720, USA}

\author{Dragomir~N.~Neshev}
\affiliation{Nonlinear Physics Center and Center for Ultrahigh Bandwidth Devices for Optical Systems (CUDOS),The Australian National University, Canberra, ACT 0200, Australia}

\author{Yuri~S.~Kivshar}
\affiliation{Nonlinear Physics Center and Center for Ultrahigh Bandwidth Devices for Optical Systems (CUDOS),The Australian National University, Canberra, ACT 0200, Australia}

\author{Xiang~Zhang}
\affiliation{NSF Nanoscale Science and Engineering Center, University of California, Berkeley, CA 94720, USA}
\affiliation{Materials Sciences Division, Lawrence Berkeley National Laboratory, Berkeley, CA 94720, USA}

\email{Sergey.Kruk@anu.edu.au}

\begin{abstract}
\bf{Strongly anisotropic media where the principal components of electric permittivity or magnetic permeability tensors have opposite signs are termed as hyperbolic media. Such media support propagating electromagnetic waves with extremely large wavevectors exhibiting unique optical properties. However in all artificial and natural optical materials studied to date, the hyperbolic dispersion originates solely from the electric response. This restricts material functionality to one polarization of light and inhibits free-space impedance matching. Such restrictions can be overcome in media having components of opposite signs for both electric and magnetic tensors. Here we present the experimental demonstration of the magnetic hyperbolic dispersion in three-dimensional metamaterials. We measure metamaterial isofrequecy contours  and reveal the topological phase transition between the elliptic and hyperbolic dispersion. In the hyperbolic regime, we demonstrate the strong enhancement of thermal emission, which becomes directional, coherent and polarized. Our findings show the possibilities for realizing efficient impedance-matched hyperbolic media for unpolarized light.}

\end{abstract}

\maketitle

The study of hyperbolic media and hyperbolic metamaterials have attracted significant attention in recent years due to their relatively simple geometry and many interesting properties, such as high density of states, all-angle negative refraction, and hyperlens imaging beyond the diffraction limit~\cite{smith2003electromagnetic, krishnamoorthy2012topological, poddubny2013hyperbolic}. Usually, both artificial~\cite{yao2008optical, noginov2009bulk, kabashin2009plasmonic,kanungo2010experimental,noginov2010controlling,Wirtz:2011:NNano} and natural~\cite{sun2014indefinite,dai2014tunable} media with hyperbolic dispersion are uniaxial materials whose axial and tangential dielectric permittivities have opposite signs. In general, however, the propagation of electromagnetic waves inside a material is defined by both the dielectric permittivity and magnetic permeability tensors. Specifically, the electric response defines the dispersion for the TM linearly polarized light, and the magnetic response defines the dispersion for TE polarization (see details in Supplementary Note 1). Therefore the ability to control both the electric permittivity and magnetic permeability gives a full flexibility for the dispersion engineering for any arbitrary polarization and nonpolarized light. This is of a major importance for the efficient interaction with randomly positioned emitters or thermal radiation. Moreover, a control over both electric and magnetic responses allows one to engineer the material dispersion and impedance independently, and, in particular, to achieve impedance matching between a hyperbolic material and the free space. Impedance matching prevents any light reflections at the interfaces and allows for efficient light coupling and extraction from the hyperbolic materials. Therefore, the development of magnetic hyperbolic materials with a simultaneous control over both dielectric permittivity and magnetic permeability tensors remains an important milestone. In particular, it is of a special interest to realize experimentally a magnetic hyperbolic material with the effective magnetic permeability tensor having principal components of the opposite signs~\cite{smith2003electromagnetic,smith2004negative}. Such a development would open new opportunities for super-resolution imaging, nanoscale optical cavities or control over the density of photon states and, in particular, the magnetic density of states for enhancing brightness of magnetic emitters~\cite{Hussain:15,PhysRevLett.114.163903}.

In recent years, we have seen the development of numerous structures with artificial magnetism.
However, these structures are largely limited to planar metasurfaces of deeply-subwavelength thickness. Importantly, many properties and functionalities of hyperbolic media rely on wave propagation inside them and therefore require essentially a three-dimensional design. For example, hyperlens super-resolution imaging relies on conversion of evanescent waves propagating in a bulk of hyperbolic media~\cite{jacob2006optical,liu2007far}. Nowadays the realization of three dimensional metamaterial structures~\cite{soukoulis2011past} is at the edge of technological possibilities associated with extreme fabrication difficulties and material constraints.  To date no photonic structures with hyperbolic dispersion in the magnetic response have been demonstrated, and such types of structures are only known for microwave systems~\cite{sun2010low,shchelokova2014magnetic}.

Here we demonstrate experimentally optical magnetic hyperbolic metamaterial with the principal components of the magnetic permeability tensor having the opposite signs. We directly observe a topological transition between the elliptic and hyperbolic dispersions in metamaterials. In the hyperbolic regime the length of wave-vectors inside the metamaterial is diverging towards infinity. We reveal the effect of the hyperbolic dispersion on thermal emission of metamaterials, where the magnetic hyperbolic metamaterial demonstrates enhanced, directional, coherent and polarized thermal emission. Our experimental observations are supported by analytical calculations as well as full-wave numerical simulations.

\section*{Results}

\subsection*{Sample fabrication}
\begin{figure*}
\centering
\includegraphics[width=0.99\textwidth]{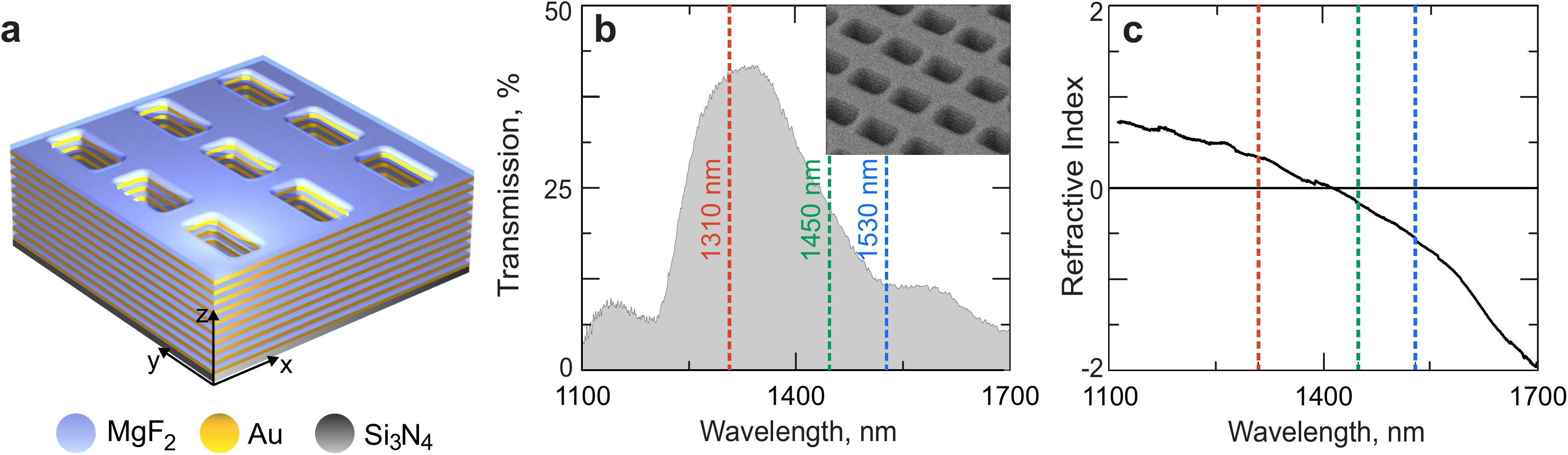}
\caption{{\bf Multilayer fishnet metamaterial.} (a) Sketch of the structure. Thicknesses of MgF$_2$ and Au layers are 45~nm and 30~nm, correspondingly. Thickness of Si$_3$N$_4$ membrane is 50 nm. Lattice period is 750$\times$750 nm. Size of holes is 260$\times$530 nm. (b) Experimentally measured transmission spectrum of the fishnet metamaterial. Inset shows a SEM image of the fabricated structure. (c) Effective refractive index of the fishnet metamaterial extracted for the normal incidence. The marked lines in (b) and (c) represent the wavelengths in the regions of elliptic dispersion (red), crossover  optical topological transition (green), and  hyperbolic dispersion (blue).}
\label{fig:3D_MMs_normal_incidence}
\end{figure*}

To realize a magnetic hyperbolic medium in optics, we employ multilayer fishnet metamaterials, known as the bulk-type metamaterials with negative refractive index at optical frequencies~\cite{valentine2008three}. Multilayer fishnets were predicted theoretically to possess a magnetic hyperbolic dispersion~\cite{kruk2012spatial}, but direct measurements of their dispersion remained out of reach. To test this, we fabricate fishnet a metamaterial by using focused ion beam milling through a stack of 20 alternating films of gold and magnesium fluoride (see details in Methods). The sample is fabricated on a 50~nm thin silicon nitride membrane. A sketch and SEM image of the fabricated structure are shown in Fig.~\ref{fig:3D_MMs_normal_incidence}. The fishnets feature high optical transmission in the near-infrared spectral region, exhibiting a transmission maximum of 42$\%$ at $\sim1320$\,nm wavelength, as shown in Fig.~\ref{fig:3D_MMs_normal_incidence}b. We also measure the fishnet refractive index at normal incidence by using spectrally and spatially resolved interferometry~\cite{o2012reflective}. The fishnet's refractive index shown in Fig.~\ref{fig:3D_MMs_normal_incidence}c is constantly decreasing with an increase of the wavelength exhibiting negative values at wavelengths above 1410\,nm.

\subsection*{Angular dispersion measurements}
To reconstruct the dispersion isofrequency contours experimentally, we determine the length of the \textbf{k}-vectors of light inside the materials for a range of different directions. For this we measure both amplitude and phase of the transmitted and reflected light and find the  \textbf{k}-vectors via the reverted Fresnel equations (see details in Supplementary Note 2).

For the phase measurements, we employ interferometry techniques. Specifically, for measuring a phase in transmission, we use the Mach-Zehnder type interferometer~\cite{saleh2007fundamentals}, while for the measurements of a phase in reflection we employ the Michelson-Morley interferometer~\cite{saleh2007fundamentals}. In order to resolve transmission and reflection at different angles, we focus and collect the light using objective lens with high numerical aperture (Olympus LCPLN100XIR NA 0.85) and project the objective's back-focal plane image onto an infrared camera (Xenics XS-1.7-320). The resulting image on a camera represents the \textbf{k}-space spectrum of the fishnet metamaterial with the central point of the image corresponding to the \textbf{k}-vectors normal to the fishnet's surface. The edge of the image corresponds to \textbf{k}-vectors oblique to the fishnet at an angle $\sim 58^\circ$, limited by the numerical aperture of the objective. We note that for transmission measurements the numerical aperture is limited to 0.7 by the finite size of the silicon nitride membrane window etched into the supporting silicon handle wafer. This restriction also assures small non-paraxial effects due to the sharp focusing. In order to obtain the phase information, we interfere the back-focal plane image with a reference beam. To reconstruct the phase information from the interference pattern we employ off-axis digital holography technique~\cite{cuche1999simultaneous} (see details in Supplementary Note 3).

\begin{figure*}
\centering
\includegraphics[width=0.8\textwidth]{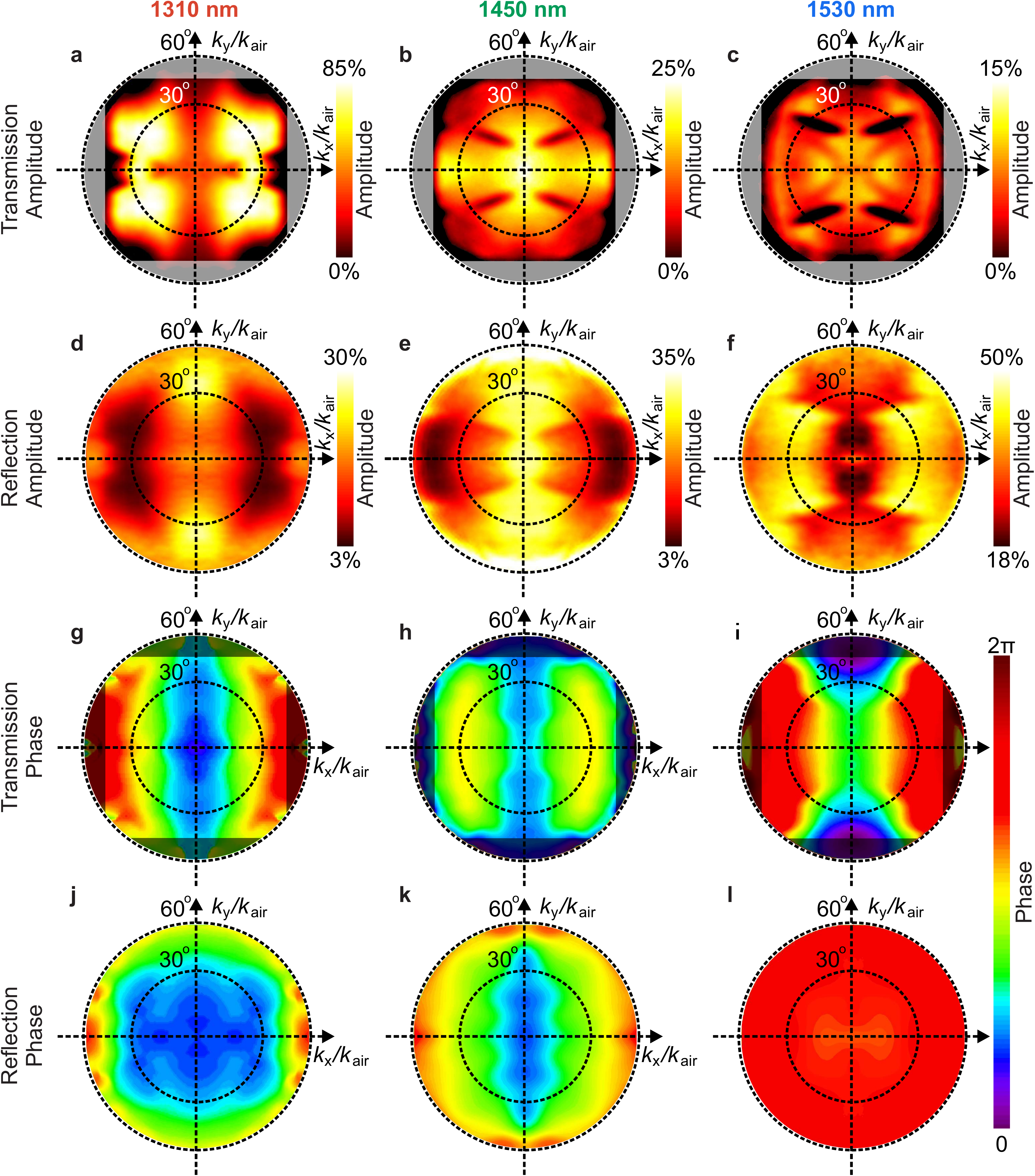}
\caption{{\bf Experimental results.} Measured (a-c) transmission and (d-f) reflection amplitudes, (g-i) transmission and (j-l) reflection phase for three different wavelengths: (a,d,g,j) 1310 nm, (b,e,h,k) 1450 nm, and (c,f,i,l) 1530 nm. All the measurements are performed for the range of incident angles $0^{\circ} - 60^{\circ}$ and are plotted versus wave-vector components $k_\text{x}$ and $k_\text{y}$ normalized by the length of the wave-vector in air $k_\text{air}$. Horizontal axes correspond to TM-polarized illumination. Vertical axes correspond to TE-illumination. Square apertures for the transmission amplitude and phase measurements show the numerical aperture limited by the size of the windows in the supporting silicon wafer.}
\label{fig:3D_MMs_fourier-interfer}
\end{figure*}

We measure the complex transmission  and reflection for three different wavelengths: 1310\,nm, 1450\,nm and 1530\,nm, marked in  Figs.~\ref{fig:3D_MMs_normal_incidence}b,c with red, green and blue, correspondingly. We notice that for the first wavelength the metamaterial exhibits positive refractive index for normal incidence of light; for the second wavelength the refractive index is close to zero and for the latter wavelength the refractive index is negative [see Fig.~\ref{fig:3D_MMs_normal_incidence}(c)]. The results of our angular measurements are presented in Fig.~\ref{fig:3D_MMs_fourier-interfer}. We use a linearly polarized light source with the electric field polarized in the $x$-direction. After the objective lens, the focusing beam has TE-polarization along the $k_\text{y}$-axis and TM-polarization along the $k_\text{x}$-axis correspondingly. As a result, the back-focal plane images (Fig.~\ref{fig:3D_MMs_fourier-interfer}) contain information about the optical response of the sample with respect to both TE- and TM-polarized light along the $k_\text{y}$ and $k_\text{x}$ axes, correspondingly. We notice that the material's magnetic dispersion is measured along the $k_\text{y}$ axis, and the electric dispersion is measured along the $k_\text{x}$ axis.

From our measurements we can see that for a specific range of angles of incidence at 1310 nm and at 1450 nm wavelengths, the reflection becomes lower than 4$\%$, which, for comparison, is lower than the reflection of glass. This is a direct consequence of the impedance matching of the metamaterial to air.

Next we analyze the phase accumulation of light inside the metamaterial [Figs.~\ref{fig:3D_MMs_fourier-interfer}(g-i)]. At 1310~nm wavelength [Fig.~\ref{fig:3D_MMs_fourier-interfer}(g)], the phase accumulation increases from the centre (normal incidence) to the edges (60$^\circ$ oblique incidence). This is similar to the response of a usual dielectric, where phase accumulation increases with a growth of an optical path inside the material. At 1450~nm wavelength, the phase accumulation along the $k_\text{y}$-axis remains nearly unchanged being close to zero for the entire range of incident angles. This corresponds to the case of $\varepsilon$-near-zero (ENZ)~\cite{alu2007epsilon} and $\mu$-near-zero (MNZ)~\cite{Engheta2006} materials.  Finally, at 1530~nm wavelength the phase accumulation is decreasing from the center to the edges along the $k_\text{y}$-axis, while it is increasing along the $k_\text{x}$-axis.

\begin{figure*}
\centering
\includegraphics[width=0.99\textwidth]{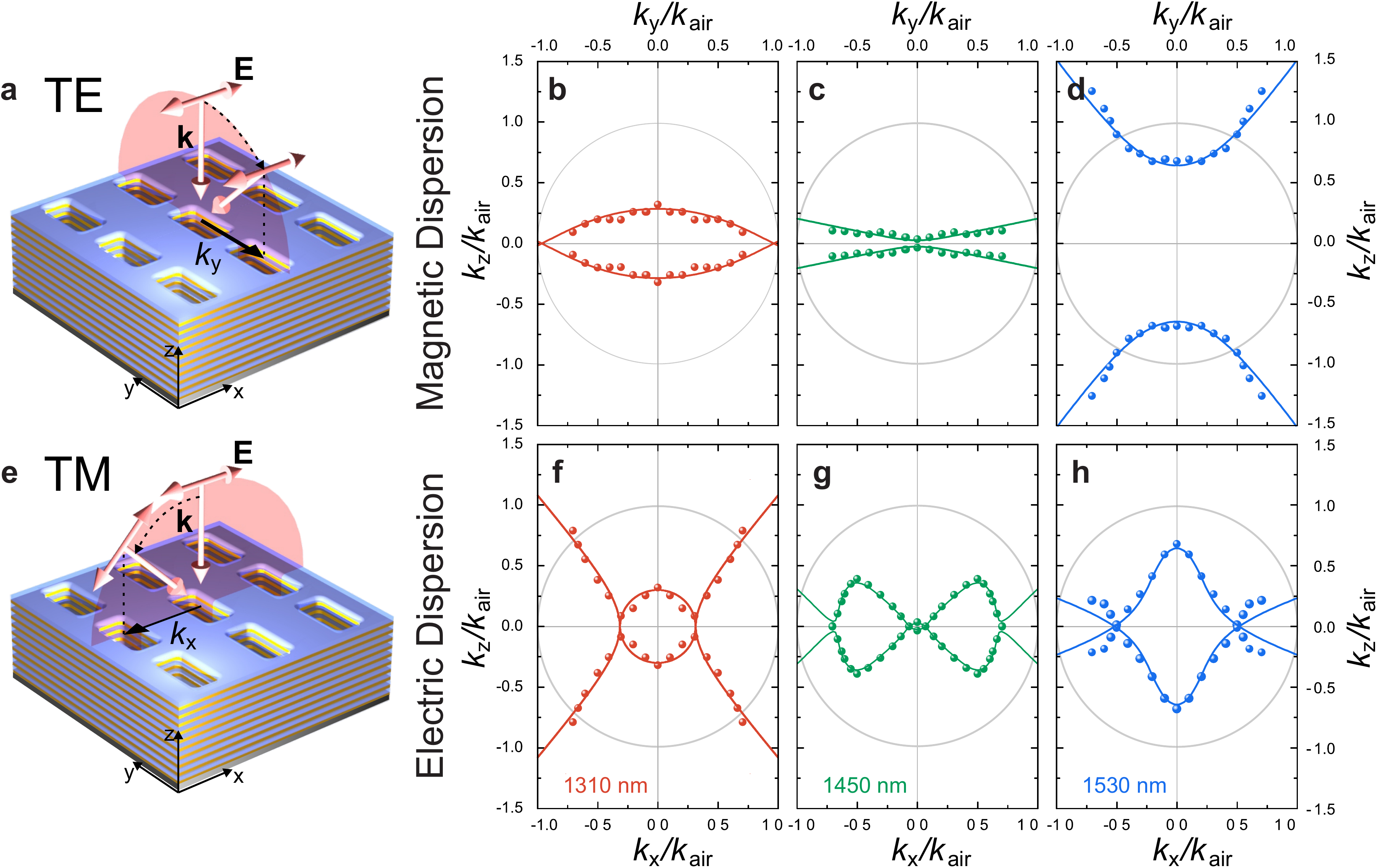}
\caption{{\bf Experimental observation of a transition from elliptic to hyperbolic dispersion.} (a) Sketch of the TE-illumination geometry showing relative orientations of the sample, wave-vector \textbf{k} and electric field \textbf{E}. (b-d) Isofrequency dispersion contours for the TE-polarization at wavelengths 1310~nm, 1450~nm, and 1530~nm, respectively. Wave-vector components $k_\text{y}$ and $k_\text{z}$ are normalized by the length of the wave-vector in air $k_\text{air}$. (e) Sketch of the TM-illumination geometry. (f-h) Isofrequency dispersion contours for the TM-polarization at wavelengths 1310~nm, 1450~nm, and 1530~nm, respectively. Dots mark experimental data, and lines correspond to analytical results. Grey circles correspond to an isofrequency contour of light in vacuum.}
\label{fig:3D_MMs_isosurfaces}
\end{figure*}

\subsection*{Isofrequency analysis}
Next, we reconstruct the isofrequency contours out of the measured transmission and reflection data (see details in Supplementary Note 2). For the three wavelengths and two polarizations these dispersion contours are shown in Figs.~\ref{fig:3D_MMs_isosurfaces}(b-d,f-h).

We use an analytical approach to reveal the shapes of isofrequecy surfaces obtained experimentally. We write explicitly a set of two equations for two principal linear polarizations: TE and TM. Without a lack of generality, we assume that for TE-polarization the electric field component is pointing in the \emph{x}-direction, and for TM-polarization the magnetic field component is pointing in the \emph{y}-direction. The resulting dispersion relations take the form:

\begin{equation}
\textrm{TE:}\quad\frac{k_\text{y}^{2}}{\varepsilon_\text{x}\mu_\text{z}}+\frac{k_\text{z}^{2}}{\varepsilon_\text{x}\mu_\text{y}}=\frac{\omega^{2}}{c^2};\qquad \textrm{TM:}\quad\frac{k_\text{x}^{2}}{\varepsilon_\text{z}\mu_\text{y}}+\frac{k_\text{z}^{2}}{\varepsilon_\text{x}\mu_\text{y}}=\frac{\omega^{2}}{c^2}.
\label{eq:3D_TE_TM}
\end{equation}

Thus the material response is described by a set of parameters $\varepsilon_\text{x}$,  $\varepsilon_\text{z}$,  $\mu_\text{y}$ and  $\mu_\text{z}$. For the case of purely real values of parameters (e.g. for materials with no loss or gain) the dispersion equations describe two types of isofrequency contours: either elliptic or hyperbolic depending on the relative signs of the parameters. In particular, opposite signs of the electric permittivity components $\varepsilon_\text{x}$ and  $\varepsilon_\text{z}$ lead to hyperbolic isofrequecy contours for TM-polarization, while the opposite signs of the magnetic permeability components $\mu_\text{y}$ and  $\mu_\text{z}$ lead to hyperbolic isofrequecy contours for the TE-polarization. Here we take into account the absorption of light in the metamaterial and consider the parameters $\varepsilon_\text{x}$,  $\varepsilon_\text{z}$,  and $\mu_\text{y}$ as complex numbers with the imaginary parts representing losses. We assume $\mu_\text{z}=1$ for all the cases, as we do not expect artificial magnetic response from the structure in the $z$-direction. Table~I summarizes the values of $\varepsilon_\text{x}$,  $\varepsilon_\text{z}$,  and $\mu_\text{y}$ used in our analytical model to describe the experimental data.

As we observe, at 1310 nm wavelength [Figs.~\ref{fig:3D_MMs_isosurfaces}(b,f)] all the material parameters have positive real parts, thus representing the cases of the elliptic dispersion. Interestingly, in both cases the shapes of the isofrequency contours deviate from elliptical. This effect comes from the imaginary parts of $\varepsilon_\text{x}$,  $\varepsilon_\text{z}$,  and $\mu_\text{y}$. It is known that finite material losses lead to a hybridization of propagating and evanescent modes~\cite{davoyan2011mode}. Importantly, in our case for the TM-polarization [see Fig.~\ref{fig:3D_MMs_isosurfaces}(f)] the hybridization leads to a new class of topology of isofrequency contours that is different from either elliptic or hyperbolic.

At 1530~nm wavelength and TE-polarization, the permeability coefficient  $\mu_\text{y}$ has a negative real part, which is opposite to the sign of the  permeability coefficient  $\mu_\text{z}=1$. Therefore, in this spectral region the material dispersion becomes magnetic hyperbolic [see Fig.~\ref{fig:3D_MMs_isosurfaces}(d)]. The branches of the hyperbola go beyond the isofrequency contour of light in air. We notice that the \textbf{k}-vectors with tangential components larger than $|k_\text{air}|$ are not accessible experimentally when the metamaterial is illuminated from free space. However, analytical extrapolation of experimental curves supports the existence of propagating waves with large wave-vectors. This is a key to achieve extraordinary optical properties of hyperbolic media, such as super-resolution imaging, nanoscale optical cavities and control over the density of photon states. For the other TM-polarization, however, all three material parameters $\varepsilon_\text{x}$,  $\varepsilon_\text{z}$, and  $\mu_\text{y}$ are simultaneously negative resulting in elliptic dispersion with a complex topology of isofrequency contours due to the presence of losses [see Fig.~\ref{fig:3D_MMs_isosurfaces}(h)].

At 1450~nm wavelength the parameters $\varepsilon_\text{x}$,  $\varepsilon_\text{z}$, and  $\mu_\text{y}$ are vanishing simultaneously, representing the regime of optical topological transition \cite{krishnamoorthy2012topological}. Around the topological transition, $\varepsilon_\text{x}$,  $\varepsilon_\text{z}$, and  $\mu_\text{y}$ change their signs due to the resonant nature of metamaterial's response.  This results in an increase of the phase velocity of light towards infinity inside the structure. Importantly, at this wavelength the structure supports propagating waves with \textbf{k}-vectors substantially smaller that the \textbf{k}-vectors in air, while all conventional optical materials support \textbf{k}-vectors larger that those in air. As local material parameters become close to zero, we expect to see strong contributions from nonlocal response of the metamaterials~\cite{gorlach2015nonlocality}. This implies that the permittivity coefficients $\varepsilon_\text{x}$ and  $\varepsilon_\text{z}$ become functions of the wave-vectror \textbf{k} (see details in Supplementary Note 4). We find that  for the magnetic (TE) dispersion of the fishnet metamaterials both local and nonlocal models result in the same dispersion relation. Therefore, we consider the electric (TM) dispersion equation only, which in the nonlocal case takes form:

\begin{align}
\textrm{TM:}&\quad\frac{k_\text{x}^{2}}{\varepsilon_\text{z}+\frac{\partial^2 \varepsilon_\text{z}}{\partial k_\text{x}^2}k_\text{x}^2+\frac{\partial^2 \varepsilon_\text{z}}{\partial k_\text{z}^2}k_\text{z}^2}+\frac{k_\text{z}^{2}}{\varepsilon_\text{x}+\frac{\partial^2 \varepsilon_\text{x}}{\partial k_\text{x}^2}k_\text{x}^2+\frac{\partial^2 \varepsilon_\text{x}}{\partial k_\text{z}^2}k_\text{z}^2}=\frac{\omega^{2}}{c^2}
\label{eq:spatial_TE_mainText}
\end{align}

We notice that $\mu_\text{y}=\left(1-\rfrac{\omega^2}{c^2}\rfrac{\partial^2\varepsilon_\text{x}}{\partial k_\text{z}^2}\right)^{-1}$~\cite{gorlach2015nonlocality}. We further neglect the nonlocal parameter $\rfrac{\partial^2 \varepsilon_\text{z}}{\partial k_\text{z}^2}$ as we find it to be of a minor importance~\cite{gorlach2015nonlocality}. Thus, for the case of 1450~nm wavelength, near the point of the optical topological transition of the metamaterial, we introduce two extra spatially dispersive terms $\rfrac{\partial^2\varepsilon_\text{x}}{\partial k_\text{x}^2}$ and $\rfrac{\partial^2\varepsilon_\text{z}}{\partial k_\text{x}^2}$ in order to describe the experimental dispersion. The values of the material parameters for the 1450 nm wavelength are also given in Table~I.  Our results suggest that a wide range of nontrivial isofrequency dispersion contours can be realized by an appropriate tuning of material's loss, gain and spatial dispersion. While all possible types of isofrequency contours for local media without loss/gain are limited to the second-order geometrical curves~\cite{Berger2009} (such as an ellipse or hyperbola), the presence of loss, gain and spatial dispersion extends the possible cases of   isofrequency contours to  the fourth-order curves. This leads to new topologies of the metamaterial dispersion.

\begin{table*}[htb]
\caption{\label{tab:symmetry}Effective parameters of the metamaterial dispersion}
\begin{tabular}[t]{|c||m{4cm}|m{4cm}|m{4cm}|}
\hline
Parameter & 1310 nm & 1450 nm * & 1530 nm\\
\hline
\hline
$\varepsilon_\text{x}$ & 0.45+i 0.8 & -0.045+i 0.03 & -0.14+i 0.25\\
\hline
$\varepsilon_\text{z}$ & 0.2+i 0.01 & 0.08+i 0.002 & -0.02+i 0.023\\
\hline
$\mu_\text{y}$ & 0.18+i 0.3 & 0.06+i 0.13 & -2+i 0.1\\
\hline
$\mu_\text{z}$ & 1 & 1 & 1\\
\hline
\hline
 \multicolumn{4}{|l|}{$*$ Nonlocal parameters at 1450~nm wavelength $\rfrac{\partial^2\varepsilon_\text{x}}{\partial k_\text{x}^2}=7$,  $\rfrac{\partial^2\varepsilon_\text{z}}{\partial k_\text{x}^2}=0.82$,  and $\rfrac{\partial^2\varepsilon_\text{z}}{\partial k_\text{z}^2}=0$} \\
 \hline
\end{tabular}
\end{table*}

In addition, we use full-wave numerical simulations to calculate material's isofrequency contours (see details in Supplementary Note 5 and Supplementary Figure 1). We find that numerical results are in a good agreement with both experimental measurements and analytical calculations.

\subsection*{Manipulation of thermal emission from fishnet metamaterials}
\begin{figure*}
\centering
\includegraphics[width=0.99\textwidth]{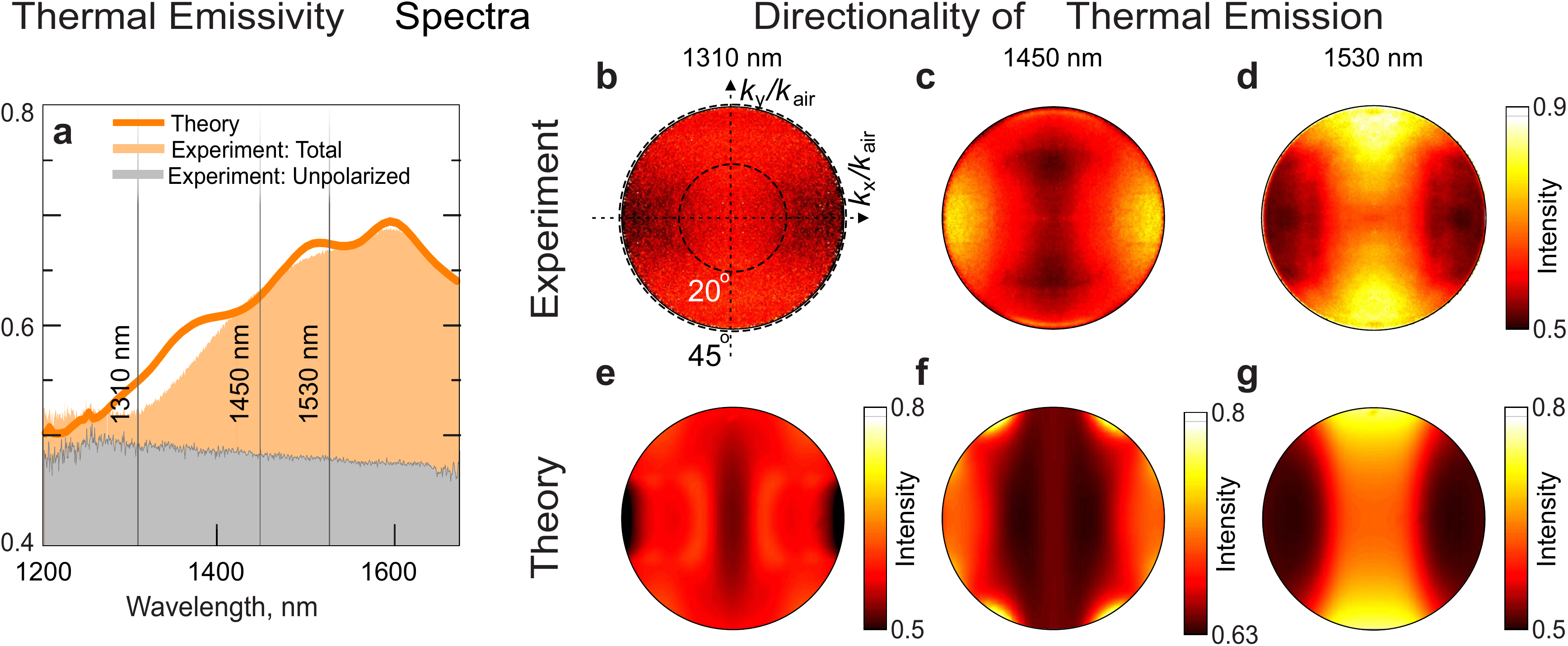}
\caption{{\bf Thermal emission from bulk magnetic-hyperbolic metamaterials.} (a) Spectrum of thermal emission normalized by the black body spectrum (emissivity). The unpolarized portion of emission is shown in gray, whereas the total emissivity is shown in orange above the unpolarized part. Line represents the corresponding theoretical calculation for the total emissivity. (b-d) Experimentally measured directionality of thermal emissivity at wavelengths 1310~nm, 1450~nm, and 1530~nm, respectively  plotted versus wave-vector components $k_\text{x}$ and $k_\text{y}$ normalized by the length of the wave-vector in air $k_\text{air}$. (e-g) Theoretically calculated directionality for the same three wavelengths, respectively. Images (c-g) have the same coordinate system as (b).}
\label{fig:3D_MMs_Thermal}
\end{figure*}

Next, we study the effect of hyperbolic dispersion on far-field thermal emission. In our experiments, we heat the sample up to 400$^\circ$C with a ceramic heater. At this temperature, the metamaterial gives relatively bright thermal emission in the spectral region of interest, and it remains undamaged by heating. We collect the thermal emission of the fishnet metamaterial sample by an objective lens with 10~mm working  distance and 0.7 numerical aperture. In our experiments, we ensure that only thermal emission from the metamaterial sample is collected by the objective lens. We then direct it onto an infrared spectrometer and measure the thermal emission spectra. We take a reference measurement of the thermal emission from a silicon sample next to the fishnet metamaterial The reference measurement allows us to find emissivity of fishnets (radiation of fishnets normalized by the black body radiation) by using the known emissivity of silicon~\cite{ravindra2001emissivity}  and, in particular, characterize the degree of polarization of the emitted light (see details in Supplementary Note 6 and Supplementary Figures 2 and 3). We then measure the polarization states of the emissivity by employing Stokes vector formalism (see details in Supplementary Note 6 and Supplementary Figures 4 and 5). We find that in the spectral region with the magnetic elliptic dispersion, the thermal emission remains largely unpolarized. However, the degree of polarized light grows rapidly as we approach the point of the optical topological transition. In the spectral region of the magnetic hyperbolic dispersion, thermal emission becomes partially linearly polarized. Figure~\ref{fig:3D_MMs_Thermal}(a) shows the emissivity spectra of our sample. We notice that unpolarized fraction of the emissivity remains almost unchanged over the measurement spectral range. The polarized part of the emission however, increases at around the topological transition region and in the region with hyperbolic dispersion. These phenomena can be explained only by the enhanced density of photon states due to the magnetic hyperbolic dispersion. In addition, we argue that our far-field results suggest that the near-field thermal radiation can be characterized as super-Planckian, i.e. exceeding the black body limit~\cite{guo2012broadband}.

Further, we study directionality of thermal emission at the three wavelengths of 1310 nm, 1450 nm and 1530 nm. For this, we translate the back-focal plane image of the collecting objective onto the infrared camera through a corresponding band-pass filter. We again employ Stokes formalism to characterize the polarization states of directionality diagrams (details of the back-focal plane polarimetry method can be found in Ref.~\cite{kruk2014spin}). With this method, we retrieve the polarized portion of the thermal emission and plot its directionality diagrams for the three wavelengths in Figs.~\ref{fig:3D_MMs_Thermal}(b-d). We notice that the directionality of emission at 1310~nm (elliptic dispersion) is not pronounced, while emission at 1530~nm has a noticeable north-south directionality. Importantly, the directions of high thermal radiation correspond to the directions with large \textbf{k}-vectors on the magnetic hyperbolic dispersion curve in Fig.~\ref{fig:3D_MMs_isosurfaces}(d). Emission at the point of topological transition exhibits noticeable directionality as well; in particular, the emission in the direction normal to the sample is suppressed [the center of the image in Fig.~\ref{fig:3D_MMs_Thermal}(c)]. This corresponds to the region with near-zero \textbf{k}-vectors. The fact that in the regime of the magnetic hyperbolic dispersion the thermal emission is directional implies that it exhibits a high degree of spatial coherence.

We further calculate the spectra and directionalities of thermal emission theoretically (see details in Supplementary Note 7). The results of our calculations are sown in Fig.~\ref{fig:3D_MMs_Thermal}(a) with a line for the spectral dencity and Figs.~\ref{fig:3D_MMs_Thermal}(e-g) for the directionality.  The calculated spectra and directionality diagrams show an excellent qualitative agreement with our experimental measurements.

In addition, we compare experimental thermal emission directionalities at 400$^\circ$C with experimentally measured absorption directionalities at room temperature (see details in Supplementary Note 6 and Supplementary Figure 6). The directionalities look similar, while resembling some differences in details associated with the change of material properties with temperature.


\section*{Discussion}
We have demonstrated experimentally optical magnetic hyperbolic metamaterial with the principal components of the  magnetic permeability tensor having the opposite signs. We have developed an experimental method for direct measurements of isofrequency dispersion contours of three-dimensional metamaterials and directly observed a topological transition between the elliptic and hyperbolic dispersions in metamaterials. In the hyperbolic regime the length of wave-vectors inside the metamaterial is diverging towards infinity.

We have applied an analytical theory that takes into account losses and  spatial dispersion to describe the measured isofrequency contours, and demonstrated the importance of nonlocal contributions~\cite{Orlov:2011:PRB.84.045424} in the regime of optical topological transitions, accociated with vanishing local parameters. A control of loss, gain and spatial nonlocalities in metamaterials opens-up new opportunities for engineering isofrequency dispersion contours beyond elliptic or hyperbolic, with non-trivial geometry and topology. The magnetic hyperbolic dispersion of metamaterials together with their electric response enables impedance matching between hyperbolic media and air, resulting in an efficient light transfer through interfaces. Our results suggest that other three-dimensional metamaterials assembled from magnetically polarizable or chiral elements~\cite{liu2008three, pendry2004chiral, gansel2009gold} may posses magnetic hyperbolic dispersion as well.

In addition, we have studied the effect of the hyperbolic dispersion on thermal emission of matematerials, and revealed that the magnetic hyperbolic metamaterial demonstrates enhanced, directional, coherent and polarized thermal emission. These results suggest an advanced thermal management that can find applications in thermophotovoltaics~\cite{lenert2014nanophotonic}, scanning thermal microscopy~\cite{de2006thermal}, coherent thermal sources~\cite{greffet2002coherent}, and other thermal devices.

\section*{Methods}

\subsection*{Nanofabrication}

The bulk fishnet metamaterial is fabricated on a suspended 50 nm low-stress silicon nitride  (Si$_3$N$_4$) membrane made from standard MEMS fabrication technologies. The metal-dielectric stack is then deposited onto Si$_3$N$_4$ membrane using layer-by-layer electron beam evaporation technique at pressure $\sim$ Torr without vacuum break. The chamber temperature is cooled down upon each layer of evaporation to avoid buildup of excessive heating and stress. Essentially ten repeating layers of gold (Au, 30 nm) and magnesium fluoride (MgF$_2$, 45 nm) are deposited. Next, the  sample is turned upside down and mounted on a special stage holder which has a matching trench that prevents any mechanical contact with the fragile multilayer structure sitting on the membrane. The nanostructures are milled by Gallium (Ga) focused ion-beam (FIB) from the membrane side. Milling from the membrane side prevents the implantation of Ga ions into the metal layers at the unpatterned areas which reduces optical losses and improves the overall quality. This is essential to mask the implantation of Ga ions into the metal layers at the unpatterned areas. The final structure has a slight sidewall angle along the thickness direction, but is found to have minor influence on the optical properties. Another important advantage of FIB fabrication of the structure on a thin membrane compared to conventional bulk substrates, is the ability for Ga ions to enter the free space (i.e. no substrate for Ga ions to accumulate and cause undesired contamination and absorption).

\section{Acknowledgements}
We thank D. Smith and D. Basov for discussions and also acknowledge useful suggestions from S. Fan and C. Simovski. The work was partially supported by the Australian Research Council.

\section*{Author Contributions}\label{sec:Author Contributions}
S.S.K., D.N.N, and Y.S.K. conceived the idea. S.S.K. conducted the experiments. E.P-S. assisted with the spatially-resolved interferometric measurements. Z.J.W., K.O. and X.Z. designed the samples. Z.J.W. fabricated the samples. Z.J.W. and K.O. conducted sample quality control. S.S.K. performed analytical modelling and numerical simulations.  S.S.K. and Y.S.K. wrote the manuscript. S.S.K., Z.J.W., K.O., D.N.N., Y.S.K. and X.Z. analyzed the experimental and theoretical results. All authors contributed to discussions and edited the manuscript.

\section*{Competing financial interests}
The authors declare no competing financial interests.

\end{document}